\documentclass[twocolumn,showpacs,amsmath,amssymb,prl]{revtex4}
\usepackage{graphicx}% Include figure files
\usepackage{dcolumn}% Align table columns on decimal point
\usepackage{bm}% bold math

\begin{document}

%\preprint{APS/123-QED}

\title{Cold Atom Clock Test of Lorentz Invariance in the Matter Sector}% Force line breaks with \\

\author{Peter Wolf}
 \altaffiliation[On leave from ]{Bureau International des Poids et Mesures, Pavillon de Breteuil, 92312 S\`evres Cedex, France.}%Lines break automatically or can be forced with \\
\author{Fr\'ederic Chapelet}%
\author{S\'ebastien Bize}%
\author{Andr\'e Clairon}%
\affiliation{LNE-SYRTE, Observatoire de Paris, 61 Av. de l'Observatoire, 75014 Paris, France}%

\date{\today}% It is always \today, today,

\begin{abstract}
We report on a new experiment that tests for a violation of Lorentz invariance (LI), by searching for a dependence of atomic transition frequencies on the orientation of the spin of the involved states (Hughes-Drever type experiment). The atomic frequencies are measured using a laser cooled $^{133}$Cs atomic fountain clock, operating on a particular combination of Zeeman substates. We analyze the results within the framework of the Lorentz violating standard model extension (SME), where our experiment is sensitive to a largely unexplored region of the SME parameter space, corresponding to first measurements of four proton parameters and improvements by 11 and 13 orders of magnitude on the determination of four others. In spite of the attained uncertainties, and of having extended the search into a new region of the SME, we still find no indication of LI violation.
\end{abstract}

\maketitle

Lorentz Invariance (LI) is the fundamental postulate of Einstein's 1905 theory of relativity, and therefore at the heart of all accepted theories of physics. It characterizes the invariance of the laws of physics in inertial frames under changes of velocity or orientation. This central role, and indications from unification theories \cite{KostoSam,Damour1,Gambini} hinting toward a possible LI violation, have motivated tremendous experimental efforts to test LI.

A comprehensive theoretical framework to describe violations of LI in all fields of present day physics has been developed over the last decade \cite{Kosto1}: the Lorentz violating Standard Model Extension (SME), motivated initially by possible Lorentz violating phenomenological effects of string theory \cite{KostoSam}. In its minimal form the SME characterizes Lorentz violation by 19 parameters in the photon sector and 44 parameters per particle \cite{KL,Bluhm} in the matter sector, of which 40 are accessible to terrestrial experiments at first order in $v_\oplus/c$ \cite{Bluhm} ($v_\oplus$ is the orbital velocity of the Earth and $c = 299792458$ m/s). Existing bounds for the proton (p$^+$), neutron (n) and electron (e$^-$) come from clock comparison and magnetometer experiments using different atomic species (see \cite{KL} and references therein, \cite{Phillips,Bear,Hou,Cane}), from resonator experiments \cite{Muller2005,Wolf2004,Muller}, and from Ives-Stilwell experiments \cite{Lane2005,Saathoff}. They are summarized in tab. \ref{SME_tab}, together with the results reported in this work.

Based on the SME analysis of numerous atomic transitions in \cite{KL,Bluhm}, we have chosen to measure a particular combination of transitions in the $^{133}$Cs atom. This provides good sensitivity to the quadrupolar SME energy shift of the proton (as defined in \cite{KL}) in the spin $7/2$ Cs nucleus, whilst being largely insensitive to magnetic perturbations (first order Zeeman effect). The corresponding region of the SME parameter space has been largely unexplored previously, with first limits on some parameters set only very recently \cite{Lane2005} by a re-analysis of the Doppler spectroscopy experiment (Ives-Stilwell experiment) of \cite{Saathoff}. Given the large improvements (11 and 13 orders of magnitude on four parameters) we obtain with respect to those results, and the qualitatively new region explored (first measurements of four parameters), our experiment had comparatively high probability for the detection of a Lorentz violating signal. However, our results clearly exclude that possibility.

\begin{table}
\caption{Orders of magnitude of present limits (in $\rm{GeV}$) on Lorentz violating parameters in the
minimal SME matter sector and corresponding references. Indices $J, K$ run over $X,Y,Z$ with $J \neq K$. Limits from the present work are in bold type, with previous limits, when available, in brackets.}
\begin{ruledtabular}
\begin{tabular}{ccccc}
Parameter & p$^+$ & n & e$^-$ & Ref.\\
\hline
$\tilde{b}_X$, $\tilde{b}_Y$ & $10^{-27}$ & $10^{-31}$ & $10^{-29}$ & \cite{Phillips}, \cite{Bear}, \cite{Hou} \\
$\tilde{b}_Z$ & - & - & $10^{-28}$ & \cite{Hou} \\
$\tilde{b}_T$, $\tilde{g}_T$, $\tilde{H}_{JT}$, $\tilde{d}_\pm$ & - &  $10^{-27}$ & - & \cite{Cane}\\
$\tilde{d}_Q$, $\tilde{d}_{XY}$, $\tilde{d}_{YZ}$ & - &  $10^{-27}$ & - & \cite{Cane}\\
$\tilde{d}_X$, $\tilde{d}_Y$ & $10^{-25}$ & $10^{-29}$ & $10^{-22}$ & \cite{KL}, \cite{Cane}, \cite{KL}\\
$\tilde{d}_{XZ}$, $\tilde{d}_{Z}$ & - & - & - \\
$\tilde{g}_{DX}$, $\tilde{g}_{DY}$ & $10^{-25}$ & $10^{-29}$ & $10^{-22}$ &\cite{KL}, \cite{Cane}, \cite{KL}\\
$\tilde{g}_{DZ}$, $\tilde{g}_{JK}$ & - & - & - \\
$\tilde{g}_{c}$ & - & $10^{-27}$ & - & \cite{Cane}\\
$\tilde{g}_{-}$, $\tilde{g}_{Q}$, $\tilde{g}_{TJ}$ & - & - & - \\
$\tilde{c}_{Q}$ & ${\bf 10}^{{\bf -22}(-11)}$ & - & $10^{-9}$ & \cite{Lane2005,Saathoff} \\
$\tilde{c}_X$, $\tilde{c}_Y$ & ${\bf 10^{-25}}$ & $10^{-25}$ & $10^{-19}$ &  \cite{KL}, \cite{Muller2005,Wolf2004,Muller} \\
$\tilde{c}_Z$, $\tilde{c}_{-}$ & ${\bf 10^{-25}}$ & $10^{-27}$ & $10^{-19}$ &  \cite{KL}, \cite{Muller2005,Wolf2004,Muller} \\
$\tilde{c}_{TJ}$ & ${\bf 10}^{{\bf -21}(-8)}$ & - & $10^{-6}$ & \cite{Lane2005,Saathoff} \\
\end{tabular}
\end{ruledtabular}
\label{SME_tab}
\end{table}

We use one of the laser cooled fountain clocks operated at the Paris observatory, the $^{133}$Cs and $^{87}$Rb double fountain FO2 (see \cite{BizeJPB} for a detailed description). We run it in Cs mode on the $|F=3\rangle \leftrightarrow |F=4\rangle$ hyperfine transition of the $6S_{1/2}$ ground state. Both hyperfine states are split into Zeeman substates $m_F=[-3,3]$ and $m_F=[-4,4]$ respectively. The clock transition used in routine operation is $|F=3,m_F=0\rangle\leftrightarrow |F=4,m_F=0\rangle$ at 9.2 GHz, which is magnetic field independent to first order. The first order magnetic field dependent Zeeman transitions ($|3,i\rangle\leftrightarrow |4,i\rangle$ with $i=\pm 1,\pm 2,\pm 3$) are used regularly for measurement and characterization of the magnetic field, necessary to correct the second order Zeeman effect of the clock transition.

A detailed description of the minimal SME as applied to the perturbation of atomic energy levels and transition frequencies can be found in \cite{KL,Bluhm}. Based on the Schmidt nuclear model \cite{Schmidtnote} one can derive the SME frequency shift of a Cs $|3,m_F\rangle\leftrightarrow |4,m_F\rangle$ transition in the form

\begin{eqnarray}
\label{clockshift}
\delta\nu &=& \frac{m_F}{14 h}\sum_{w=p,e}\left(\beta_w\tilde{b}_3^w - \delta_w\tilde{d}_3^w + \kappa_w\tilde{g}_d^w\right)-\frac{m_F^2}{14 h}\left(\gamma_p\tilde{c}_q^p\right) \nonumber\\
&+& m_F K_Z^{(1)}B+\left(1-\frac{m_F^2}{16}\right)K_Z^{(2)}B^2
\end{eqnarray}
for the quantization magnetic field in the negative z direction (vertically downward) in the lab frame. The first two terms in (\ref{clockshift}) are Lorentz violating SME frequency shifts, the last two describe the first and second order Zeeman frequency shift (we neglect $B^3$ and higher order terms). The tilde quantities are linear combinations of the SME matter sector parameters of tab. \ref{SME_tab} in the lab frame, with p,e standing for the proton and electron respectively. The quantities $\beta_w, \delta_w, \kappa_w, \gamma_w, \lambda_w$ depend on the nuclear and electronic structure, they are given in tab. II of \cite{Bluhm}, $h$ is Planck's constant, $B$ is the magnetic field seen by the atom, $K_Z^{(1)} = 7.0084$~Hz~nT$^{-1}$ is the first order Zeeman coefficient, $K_Z^{(2)}=427.45\times 10^8$ Hz~T$^{-2}$ is the second order coefficient \cite{VanAud}. The tilde quantities in (\ref{clockshift}) are time varying due to the motion of the lab frame (and hence the quantization field) in a cosmological frame, inducing modulations of the frequency shift at sidereal and semi-sidereal frequencies, which can be searched for.

From (\ref{clockshift}) we note that the $m_F=0$ clock transition is insensitive to Lorentz violation or the first order Zeeman shift, while the Zeeman transitions ($m_F \neq 0$) are sensitive to both. Hence, a direct measurement of a Zeeman transition with respect to the clock transition allows a LI test. However, such an experiment would be severely limited by the strong dependence of the Zeeman transition frequency on $B$, and its diurnal and semi-diurnal variations. To avoid such a limitation, we "simultaneously" (see below) measure the $m_F=3$, $m_F=-3$ and $m_F=0$ transitions and form the combined observable $\nu_{c} \equiv \nu_{+3}+\nu_{-3}-2\nu_{0}$. From (\ref{clockshift})

\begin{equation}
\label{obsshift}
\nu_{c}= \frac{1}{7 h}K_p\tilde{c}_q^p-\frac{9}{8}K_{Z}^{(2)}B^2
\end{equation}
where we have used $\gamma_p = -K_p/9$ from \cite{Bluhm} ($K_p\sim 10^{-2}$ in the Schmidt nuclear model). This observable is insensitive to the first order Zeeman shift, and should be zero up to the second order Zeeman correction and a possible Lorentz violating shift in the first term of (\ref{obsshift}).  

The lab frame parameter $\tilde{c}_q^p$ can be related to the conventional sun-centered frame parameters of the SME (the parameters of tab. \ref{SME_tab}) by a time dependent boost and rotation (see \cite{Bluhm} for details). This leads to a general expression for the observable $\nu_c$ of the form

\begin{eqnarray}
\label{model}
\nu_{c}=A~&+&C_{\omega_\oplus}{\rm cos}(\omega_\oplus T_\oplus)+S_{\omega_\oplus}{\rm sin}(\omega_\oplus T_\oplus) \\
&+&C_{2\omega_\oplus}{\rm cos}(2\omega_\oplus T_\oplus)+S_{2\omega_\oplus}{\rm sin}(2\omega_\oplus T_\oplus), \nonumber
\end{eqnarray}
where $\omega_\oplus$ is the frequency of rotation of the Earth, $T_\oplus$ is time since 30 March 2005 11h 19min 25s UTC (consistent with the definitions in \cite{KM}), and with $A$, $C_{\omega_\oplus}$, $S_{\omega_\oplus}$, $C_{2\omega_\oplus}$ and $S_{2\omega_\oplus}$ given in tab. \ref{CS_tab} as functions of the sun frame SME parameters. A least squares fit of (\ref{model}) to our data provides the measured values given in tab. \ref{CS_tab}, and the corresponding determination of the SME parameters.

\begin{table*}
\caption{\label{CS_tab} Coefficients of (\ref{model}) to first order in $\beta \equiv v_\oplus/c$, where $\Omega_\oplus$ is the angular frequency of the Earth's orbital motion, $T$ is time since the March equinox, $\chi = 41.2^\circ$ is the colatitude of our lab, and $\eta=23.3^\circ$ is the inclination of the Earth's orbit. The measured values (in ${\rm mHz}$) are shown together with the statistical (first bracket) and systematic (second bracket) uncertainties.}
\begin{ruledtabular}
\begin{tabular}{ccc}
$A$ & $\frac{K_p}{28h}\left(1+3{\rm cos}(2\chi)\right)\left(\tilde{c}_Q+\beta\left({\rm sin}(\Omega_\oplus T)\tilde{c}_{TX}+{\rm cos}(\Omega_\oplus T)\left(2{\rm sin}\eta~\tilde{c}_{TZ}-{\rm cos}\eta~\tilde{c}_{TY}\right)\right)\right)-\frac{9}{8}K_{Z}^{(2)}B^2$ & -5.3(0.04)(25) \vspace{1mm}\\
$C_{\omega_\oplus}$ & $-\frac{3K_p}{14h}{\rm sin}(2\chi)\left(\tilde{c}_Y+\beta\left({\rm sin}(\Omega_\oplus T)\tilde{c}_{TZ}-{\rm cos}(\Omega_\oplus T){\rm sin}\eta~\tilde{c}_{TX}\right)\right)$ & 0.1(0.06)(0.03) \vspace{1mm}\\
$S_{\omega_\oplus}$ & $-\frac{3K_p}{14h}{\rm sin}(2\chi)\left(\tilde{c}_X-\beta~{\rm cos}(\Omega_\oplus T)\left({\rm sin}\eta~\tilde{c}_{TY}+{\rm cos}\eta~\tilde{c}_{TZ}\right)\right)$ & -0.03(0.06)(0.03) \vspace{1mm}\\
$C_{2\omega_\oplus}$ & $-\frac{3K_p}{14h}{\rm sin}^2\chi\left(\tilde{c}_-+\beta\left({\rm sin}(\Omega_\oplus T)\tilde{c}_{TX}+{\rm cos}(\Omega_\oplus T){\rm cos}\eta~\tilde{c}_{TY}\right)\right)$ & 0.04(0.06)(0.03) \vspace{1mm}\\
$S_{2\omega_\oplus}$ & $-\frac{3K_p}{14h}{\rm sin}^2\chi\left(\tilde{c}_Z+\beta\left({\rm sin}(\Omega_\oplus T)\tilde{c}_{TY}-{\rm cos}(\Omega_\oplus T){\rm cos}\eta~\tilde{c}_{TX}\right)\right)$ & 0.03(0.06)(0.03) \\
\end{tabular}
\end{ruledtabular}
\end{table*}

\begin{figure}[b]
\begin{center}
\includegraphics[width=7cm]{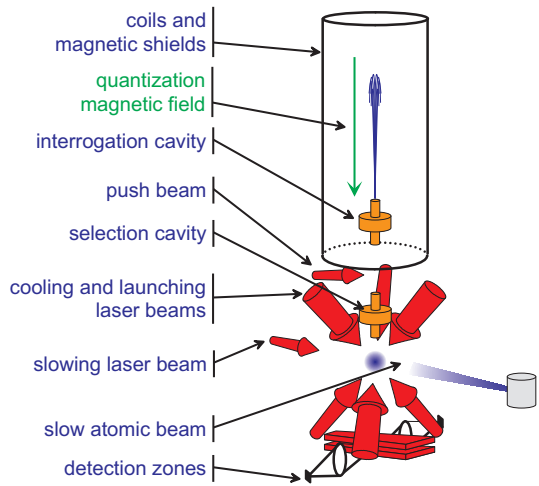}
\end{center}
\caption{Schematic view of an atomic fountain.} \label{fig:fountain}
\end{figure}

The FO2 setup is sketched in fig. \ref{fig:fountain}. Cs atoms
effusing from an oven are slowed using a chirped counter propagating laser
beam and captured in a lin $\perp$ lin optical molasses. Atoms are
cooled by six laser beams supplied by pre adjusted optical fiber couplers
precisely attached to the vacuum tank. Compared to typical FO2 clock operation \cite{BizeJPB}, the number of
atoms loaded in the optical molasses has been reduced to $\sim 2\times
10^{7}$ atoms captured in 30~ms. This reduces the collisional frequency shift of $\nu_c$ to below 0.1~mHz, and even less ($< 1 \mu$Hz) for its variation at $\omega_\oplus$ and $2\omega_\oplus$. 

Atoms are launched upwards at 3.94~m.s$^{-1}$ by using a moving
optical molasses and cooled to $\sim 1~\mu$K in the moving frame by
adiabatically decreasing the laser intensity and increasing the
laser detuning. Atoms are then selected by means of a microwave
excitation in the selection cavity performed in a bias magnetic
field of $\sim 20$~$\mu$T, and of a push laser beam. Any of the
$|3,m_F\rangle$ states can be prepared with a high
degree of purity (few $10^{-3}$) by tuning the selection microwave
frequency. 52~cm above the capture zone, a cylindrical copper cavity
(TE$_{011}$ mode) is used to probe the
$|3,m_F\rangle\leftrightarrow
|4,m_F\rangle$ hyperfine transition at 9.2~GHz. The
Ramsey interrogation method is performed by letting the atomic cloud
interact with the microwave field a first time on the way up and a
second time on the way down. After the interrogation, the
populations $N_{F=4}$ and $N_{F=3}$ of the two hyperfine levels are
measured by laser induced fluorescence, leading to a determination
of the transition probability. From the transition probability, measured on both sides of the
central Ramsey fringe, we compute an error signal to lock the
microwave interrogation frequency to the atomic transition using a
digital servo loop. The frequency corrections are applied to a
high resolution DDS synthesizer in the microwave
generator and used to measure the atomic
transition frequency with respect to the local reference.

The homogeneity and the stability of the magnetic field in the
interrogation region is a crucial point for the experiment. A
magnetic field of $203$~nT is produced by a main solenoid (length
815~mm, diameter 220~mm) and a set of 4 compensation coils, surrounded by 5 cylindrical
magnetic shields. Furthermore, magnetic field fluctuations are actively stabilized by acting on 4 additional hexagonal coils. The magnetic field in the
interrogation region is probed using the
$|3,1\rangle\leftrightarrow
|4,1\rangle$ atomic transition. Measurements of the transition frequency
as a function of the launch height show a peak to peak spatial
dependence of $230$~pT over a range of 320~mm above the interrogation
cavity with a variation of $\leq$ 0.1 pT/mm around the apogee of the atomic trajectories. Measurements of the same transition as a function of time at
the launch height of 791~mm show a magnetic field instability near
2~pT$/\sqrt{\tau}$. The long term behavior exhibits residual variations of the magnetic
field ($\sim$ 0.7~pT at $\tau=$10000~s) induced by temperature fluctuations.

The experimental sequence is tailored to circumvent the limitation
that the long term magnetic field fluctuations could cause. First
$|3,-3\rangle$ atoms are selected and the
$|3,-3\rangle\leftrightarrow
|4,-3\rangle$ transition is probed at half maximum
on the red side of the resonance (0.528~Hz below the resonance
center). The next fountain cycle, $|3,+3\rangle$
atoms are selected and the
$|3,+3\rangle\leftrightarrow
|4,+3\rangle$ transition is also probed at half
maximum on the red side of the resonance. The third and fourth fountain cycles,
the same two transitions are probed on the blue side of the resonances (0.528~Hz above the resonance
centers). This 4180~ms long sequence is repeated so as to
implement two interleaved digital servo loops finding the line
centers of both the $|3,-3\rangle\leftrightarrow
|4,-3\rangle$ and the
$|3,+3\rangle\leftrightarrow
|4,+3\rangle$ transitions. Every 400 fountain cycles, the above sequence is interrupted and the
regular clock transition
$|3,0\rangle\leftrightarrow
|4,0\rangle$ is measured for 10~s allowing for an
absolute calibration of the local frequency reference with a
suitable statistical uncertainty. Using this sequence, magnetic field fluctuations over timescales $\geq 4$~s are rejected in the combined observable $\nu_c$ and the stability is dominated by the short term
($\tau < 4$~s) magnetic field fluctuations.

\begin{figure}[b]
\begin{center}
\includegraphics[width=8cm]{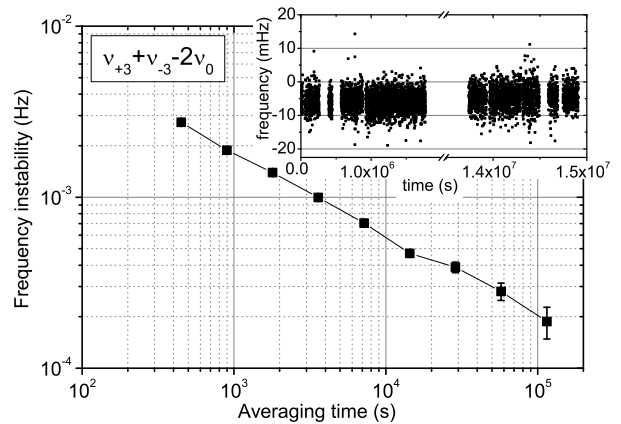}
\end{center}
\caption{Frequency stability (Alan deviation) of a $\sim 10$ day continuous stretch of measurements of $\nu_c$. The inset shows the raw data as a function of time since 30 March 2005 17h 39min 18s UTC.} \label{SMEclockPRL1}
\end{figure}

We have taken two data sets implementing the experimental sequence described above (21 days in April 2005 and 14 days in September 2005). The complete raw data (no post-treatment) is shown in fig. \ref{SMEclockPRL1}, each point representing a $\sim$~432~s measurement sequence of $\nu_{+3}+\nu_{-3}-2\nu_{0}$ as described above. Fig. \ref{SMEclockPRL1} also shows the frequency stability of a 10 day continuous stretch of data in the March set. We note the essentially white noise behavior of the data, indicating that the experimental sequence successfully rejects all long term variations of the magnetic field or of other perturbing effects. A least squares fit of the model (\ref{model}) to the complete data provides the 5 coefficients and statistical uncertainties given in tab. \ref{CS_tab}.

We note a statistically significant offset of the data from zero ($-5.3(0.04)$~mHz). This is partly due to the second order Zeeman shift (second term in (\ref{obsshift})) which amounts to $-2.0$~mHz. The remaining $-3.3$~mHz are either due to a systematic effect or indicate a genuine Lorentz violating signal.

The dominant systematic effect in our experiment is a residual first order Zeeman shift. This arises when the $m_F=+3$ and $-3$ atoms have slightly different trajectories in the presence of a magnetic field gradient. The result is a difference in first order Zeeman shift and hence incomplete cancelation in the combined observable $\nu_c$. The presence of this effect in our experiment is confirmed by the measured times of flight (TOF), which show a systematic difference of $\sim 158~\mu$s between the centers of the $m_F=\pm 3$ atomic clouds, which corresponds to a vertical separation of $\leq 623~\mu$m at apogee. Using this in a Monte Carlo (MC) simulation with the measured vertical and horizontal ($\sim 6$~pT/mm) magnetic field gradients we obtain a total offset in $\nu_c$ of $\sim 25$~mHz, assuming that the horizontal separation between the $m_F=\pm 3$ atoms is identical to the measured vertical one. We consider this to be an upper limit (the horizontal separation is likely to be less than the vertical one due to the absence of gravity), and take it as the systematic uncertainty on the determined offset ($A$ in tab. \ref{CS_tab}). To obtain the corresponding systematic uncertainty on the sidereal and semi-sidereal modulations of $\nu_c$ we have fitted the model (3) to the $m_F=\pm 3$ TOF difference. We find no significant amplitudes, so we take the uncertainties of the least squares fit as the maximum value of the modulations, leading to the systematic uncertainties on the amplitudes in tab. \ref{CS_tab}.  

Finally we use the five measurements and the relations in tab. \ref{CS_tab} to determine the values of the eight SME parameters. In doing so, we assume that there is no correlation between the three $\tilde{c}_{TJ}$ parameters and the other five parameters, and determine them independently. The results are given in tab. \ref{results}.

\begin{table}
\caption{\label{results}Results for SME Lorentz violating parameters $\tilde{c}$ for the proton, in ${\rm GeV}$ and with $J=X,Y,Z$.}
\begin{ruledtabular}
\begin{tabular}{l}
$\tilde{c}_{Q\ }=-0.3(2.2)$ $\times 10^{-22}$ \hspace{11mm} $\tilde{c}_{-}=-1.8(2.8)$ $\times 10^{-25}$ \\
$\tilde{c}_{J\ }=\hspace{6mm} \left(0.6(1.2),\hspace{3mm} -1.9(1.2),\hspace{3mm} -1.4(2.8)\right)$\hspace{5mm} $\times 10^{-25}$  \\
$\tilde{c}_{TJ}=\hspace{5mm}\left(-2.7(3.0),\hspace{3mm} -0.2(3.0),\hspace{2mm} -0.4(2.0)\right)$\hspace{3.5mm} $\times 10^{-21}$\\
\end{tabular}
\end{ruledtabular}
\end{table}

In conclusion, we have carried out a test of Lorentz invariance in the matter sector of the minimal SME using Zeeman transition in a cold $^{133}$Cs atomic fountain clock. From our data and extensive analysis of systematic effects we see no indication of a violation of LI at the present level of experimental uncertainty. Using a particular combination of the different atomic transitions we have set first limits on four proton SME parameters and improved previous limits \cite{Lane2005} on four others by 11 and 13 orders of magnitude.

Continuing the experiment regularly over a year or more will allow statistical decorrelation of the three $\tilde{c}_{TJ}$ parameters from the other five, due to their modulation at the annual frequency ($\Omega_\oplus T$ terms in tab. \ref{CS_tab}). Further improvements, and new measurements, could come from the unique capability of our fountain clock to run on $^{87}$Rb and $^{133}$Cs simultaneously. The different sensitivity of the two atomic species to Lorentz violation (see \cite{Bluhm}) and magnetic fields, should allow a measurement of all SME parameters in (\ref{clockshift}) in spite of the presence of the first order Zeeman effect. Ultimately, space clocks, like the planned ACES mission \cite{ACES} will provide the possibility of carrying out similar experiments but with faster (90 min orbital period) modulation of the putative Lorentz violating signal, and correspondingly faster data integration.


\begin{thebibliography}{99}
\bibitem{KostoSam} Kosteleck\'y V.A., Samuel S., Phys.Rev.{\bf D39}, 683, (1989).
\bibitem{Damour1} Damour T., gr-qc/9711060 (1997).
\bibitem{Gambini} Gambini R., Pullin J., Phys. Rev. {\bf D59}, 124021, (1999).
\bibitem{Kosto1} Colladay D., Kosteleck\'y V.A., Phys.Rev.{\bf D55}, 6760, (1997); Colladay D., Kosteleck\'y V.A., Phys.Rev.{\bf D58}, 116002, (1998); Kosteleck\'y V.A., Phys.Rev.{\bf D69}, 105009, (2004).
\bibitem{KL} Kosteleck\'y V.A., Lane C.D., Phys.Rev.{\bf D60}, 116010, (1999).
\bibitem{Bluhm} Bluhm R., et al., Phys. Rev. {\bf D68}, 125008, (2003).
\bibitem{Phillips} Berglund C.J., et al., Phys. Rev. Lett. {\bf 75}, 1879, (1995); Phillips D.F. et al., Phys. Rev. {\bf D63}, 111101(R), (2001); Humphrey M.A., arXiv:physics/0103068; Phys. Rev. {\bf A62}, 063405, (2000).
\bibitem{Bear} Bear D. et al., Phys. Rev. Lett. {\bf 85}, 5038, (2000).
\bibitem{Hou} Hou L.-S., Ni W.-T., Li Y.-C.M., Phys. Rev. Lett. {\bf 90}, 201101, (2003); Bluhm R. and Kostelecky V.A., Phys. Rev. Lett. {\bf 84}, 1381, (2000).
\bibitem{Cane} Can\'e F. et al., Phys. Rev. Lett. {\bf 93}, 230801, (2004).
\bibitem{Muller2005} M\"uller H., Phys. Rev. {\bf D71}, 045004, (2005).
\bibitem{Wolf2004} Wolf P., et al., Phys. Rev. {\bf D70}, 051902(R), (2004).
\bibitem{Muller} M\"uller H. et al., Phys. Rev. Lett. {\bf 91}, 2, 020401, (2003).
\bibitem{Lane2005} Lane C.D., Phys. Rev. {\bf D72}, 016005, (2005).
\bibitem{Saathoff} Saathoff G., et al., Phys. Rev. Lett. {\bf 91}, 19, 190403, (2003).
\bibitem{BizeJPB} Bize S., et al., J. Phys. {\bf B38}, S449–S468, (2005).
\bibitem{Schmidtnote} As discussed in \cite{KL} the Schmidt nuclear model only allows an approximate calculation of the SME frequency shift, with more complex models generally leading to dependences on additional SME parameters. Nonetheless, the model is sufficient to derive the leading order terms and has been used for the analysis of most experiments providing the bounds in Tab. \ref{SME_tab}.
\bibitem{VanAud} Vanier J., Audoin C., {\it The Qunatum Physics of Atomic Frequency Standards}, Adam Hilger, (1989).
\bibitem{KM} Kostelecky A.V. and Mewes M., Phys. Rev. {\bf D66}, 056005, (2002).
\bibitem{ACES} Salomon C., et al., C.R. Acad. Sci. Paris {\bf 2}, 1313, (2001).


\end{thebibliography}
\end{document}